\documentclass[a4paper,11pt]{article}
 
\newtheorem{theo}{Theorem}[section]
\newtheorem{ex}[theo]{Example}

\newtheorem{rema}[theo]{Remark}

\usepackage{amscd}
\usepackage{amssymb}
\usepackage{amsmath}

\newcommand{\nc}{\newcommand}
\nc{\beq}{\begin{equation}}
\nc{\eeq}{\end{equation}}
\nc{\beqa}{\begin{eqnarray}}
\nc{\eeqa}{\end{eqnarray}}

\begin{document}
\title{Fractional Supersymmetry and $F-$fold Lie Superalgebras}

 \author{M. Rausch de Traubenberg
\thanks{rausch@lpt1.u-strasbg.fr}$\,\,$${}^{a}$ and
M.J. Slupinski \thanks{slupins@irmasrv2.u-strasbg.fr}$\,\,$${}^{b}$\\
\\
{\small ${}^{a}${\it
Laboratoire de Physique Th\'eorique, 3 rue de
l'Universit\'e, 67084 Strasbourg, France}}\\
{\small ${}^b${\it Institut de Recherches en Math\'ematique Avanc\'ee}}\\
{\small { \it Universit\'e Louis-Pasteur, and CNRS}}\\
{\small {\it 7 rue R. Descartes, 67084 Strasbourg Cedex, France}}}        
\date{}   
\baselineskip=15pt
\maketitle


\begin{abstract}
We give infinite dimensional and finite dimensional
examples of $F-$fold Lie superalgebras. The finite dimensional
examples are obtained by an inductive procedure from Lie algebras
and Lie superalgebras.
\end{abstract}

\section{Introduction}
It is generally held that supersymmetry is the only
non-trivial extension of the Poincar\'e algebra. 
This point of view  is based on the  two theorems
\cite{cm, hls}. However, as usual, if some of the assumptions
of these two {\it no-go} theorems are relaxed symmetry beyond
supersymmetry can be constructed  \cite{ker, luis, para,
fsusy, fsusy1d,fr, qfsusy, hek, asm,am, prs, fsusy2d, ad, kk, fvir, fsusy3d, 
brs, fsusyh, flie, poly, infty}. In  all these possible extensions of 
the Poincar\'e
symmetry, new generators  are introduced. The basic structure
underlying these extensions is related to algebraic 
structures which are neither
Lie algebras, nor Lie superalgebras. In this contribution we would
like to give some results concerning fractional supersymmetry (FSUSY) 
\cite{fsusy, fsusy1d,fr, qfsusy, hek, asm,am, prs, fsusy2d, ad, kk, fvir, 
fsusy3d, brs, fsusyh, flie, poly, infty},
one of the possible extensions of supersymmetry,
and the associated algebraic structure,  
the so-called $F-$Lie algebra \cite{flie}.
Such a structure admits a $\mathbb Z_F$ grading, the zero-graded part
defining a Lie algebra, and an $F-$fold symmetric product (playing the role of
the anticommutator in the case $F=2$) allows one to express the zero graded 
 part in terms of  generators of the non zero graded part. 
In section 2 the basic definition  of $F-$ Lie  algebras will
be given. In section 3, some examples of infinite dimensional 
$F-$ Lie algebras will  be explicitly  constructed. In section  4,
some examples of finite dimensional $F-$ Lie algebras will be explicitly
constructed  with the classification of the usual Lie (super)algebras as
a guideline.

\section{$F-$Lie algebras}

The natural mathematical  structure, generalizing the concept of 
Lie superalgebras and relevant for the algebraic description of fractional
supersymmetry was introduced in \cite{flie}  and  called 
an $F-$Lie algebra. We do not 
want  to go into the  detailed definition of this structure here
and will  only recall the basic points, useful for our purpose. 
More details can be found in \cite{flie}.\\

Let $F$ be a positive integer and $q=e^{2i \frac{\pi}{F}}$.
 We consider  now a complex vector space $S$ which has an automorphism
$\varepsilon$ satisfying $\varepsilon^F=1$.
We set ${  A}_k=S_{q^k}, 1 \le k \le F-1$ and ${  B}=S_1$
($S_{q^k}$ is the eigenspace corresponding to the eigenvalue $q^k$ of
$\varepsilon$). Hence,

\[
S= B \oplus A_1 \oplus \cdots \oplus A_{F-1}.
\]
\noindent
We say that $S$ is an $F-$Lie algebra if:

\begin{enumerate}
\item $B$, the zero graded  part of $S$,   is a Lie algebra.
\item $A_i (i=1,\cdots, F-1)$,  the $i$ graded part of $S$,  
is a   representation of $B$. 
\item  There are symmetric multilinear $B-$equivariant maps 

\[
\left\{~~, \cdots,~~ \right\}:  {\cal S}^F\left(A_k\right)
\rightarrow B.
\]

\noindent
In other words, we  assume that some of the elements of the Lie algebra $B$ can
be expressed as $F-$th order symmetric products of
``more fundamental generators''.
It is easy to see that

\[
\left\{\varepsilon(a_1), \cdots, \varepsilon(a_F)\right\}=
\varepsilon\left(\left\{a_1, \cdots, a_F\right\}\right),
\forall a_1, \cdots, a_F \in { A}_k.
\]           
\end{enumerate}

\noindent
The generators of $S$ are assumed to satisfy  Jacobi identities
($b_i \in B, a_i \in A_k, 1 \le k \le F-1$):

\begin{eqnarray}
\label{eq:jac}
&&\left[\left[b_1,b_2\right],b_3\right] +
\left[\left[b_2,b_3\right],b_1\right] +
\left[\left[b_3,b_1\right],b_2\right] =0 \nonumber \\
&&\left[\left[b_1,b_2\right],a_3\right] +
\left[\left[b_2,a_3\right],b_1\right] +
\left[\left[a_3,b_1\right],b_2\right]  =0  \\
&&\left[b,\left\{a_1,\dots,a_F\right\}\right] =
\left\{\left[b,a_1 \right],\dots,a_F\right\}  +
\dots +
\left\{a_1,\dots,\left[b,a_F\right] \right\} \nonumber \\
&&\sum\limits_{i=1}^{F+1} \left[ a_i,\left\{a_1,\dots,
a_{i-1},
a_{i+1},\dots,a_{F+1}\right\} \right] =0. \nonumber
\end{eqnarray}

\noindent
The  first three identities  are consequences of the previously
defined properties but the fourth is an extra constraint.

More details  (unitarity,  representations, {\it etc.})
can be found in \cite{flie}. Let us first note that no relation between 
different graded sectors is  postulated. 
Secondly, the sub-space  $B \oplus A_k \subset S \ (k=1, \cdots, F-1)$ 
is itself an
$F-$Lie algebra. From now on, $F-$Lie algebras of the types $B \oplus A_k $
will be considered.

\section{Examples of infinite dimensional $F-$Lie algebras}

It is possible   to construct 
an $F-$Lie algebra  starting from a Lie algebra 
$\mathfrak{g}$ 
and a $\mathfrak{g}-$module ${\cal D}$. The basic idea is the following. 
We consider  $\mathfrak{g}$ a  semi-simple Lie algebra 
of rank $r$.
Let $\mathfrak{h}$ be a Cartan sub-algebra of $\mathfrak{g}$, 
let $\Phi \subset \mathfrak{h}^\star$ 
be the corresponding set of roots and let
$\mathfrak{f}_\alpha$ be the one-dimensional root space associated to 
$\alpha \in \Phi$.
We choose a basis  $\{H_i, i=1, \cdots,  r\}$ of $\mathfrak{h}$ and elements 
$E^\alpha \in \mathfrak{f}_\alpha$ such that the commutation relations  become

\begin{eqnarray}
\label{eq:lie}
\big[H_i,H_j \big] &=& 0 \nonumber \\
\big[H_i,E^\alpha \big] &=&  \alpha^i E^\alpha \\
\big[E^\alpha, E^\beta\big] &=& \left \{
\begin{array}{ll}
\epsilon\{\alpha,\beta\} E^{\alpha+\beta}& {\mathrm {~~if~~}} 
\alpha + \beta \in \Phi \cr
\frac{2\alpha.H} { \alpha.\alpha}& {\mathrm {~~if~~} } \alpha+\beta=0 \cr
0& {\mathrm {~~otherwise }}
\end{array}
\right. \nonumber
\end{eqnarray}

We now introduce $\{\alpha_{(1)},\cdots, \alpha_{(r)}\}$ 
a basis of simple roots.  
The weight lattice $\Lambda_W(\mathfrak{g}) \subset \mathfrak{h}^\star$ 
is the set of vectors
$\mu$ such that $ \frac{2 \alpha.\mu}{ \alpha.\alpha} \in \mathbb Z$ and, as is
well known, there is a  basis of the weight lattice consisting of the
fundamental weights $\{\mu_{(1)}, \cdots, \mu_{(r)} \}$ defined by
$ \frac{2 \mu_{(i)}.\alpha_{(j)}}{ \alpha_{(j)}.\alpha_{(j)}}= \delta_{ij}$.
A weight $\mu = \sum \limits_{i=1}^r n_i \mu_{(i)}$ is called dominant if
all the $n_i \ge 0$ and it is well known that the set of dominant weights
is in one to one correspondence with the set of (equivalence classes of)
irreducible finite dimensional representations of $\mathfrak{g}$.

Recall briefly how one can associate a  representation of $\mathfrak{g}$ to 
$\mu \in {\mathfrak h}^\star, \mu = \sum_{i=1}^r n_i \mu_i,
n_i \in \mathbb C$. We start with a vacuum $|\mu>$ such that

\begin{eqnarray}
\label{eq:hw}
&&E^\alpha | \mu > =0, \alpha >0,  \nonumber  \\
&&2\frac{\alpha_{(i)}.H}{\alpha^2} | \mu> = n_i | \mu>), i=1,\cdots,r.
\end{eqnarray}
\noindent
The space obtained from $| \mu >$ by  the action of 
elements of $\mathfrak{g}$: 
\[
{\cal V}_\mu= \left\{E^{-\alpha_{(i_1)}} \cdots E^{-\alpha_{(i_k)}} 
| \mu>, \alpha_{(i_1)}, \cdots, \alpha_{(i_k)} >0 \right\},
\]
\noindent
clearly defines a representation of $\mathfrak{g}$. Taking the quotient
of ${\cal V}_\mu$ by its  maximal $\mathfrak{g}-$stable subspace,
the  representation ${\cal D}_\mu$ of highest weight $\mu$  is obtained.
If the $n_i$ are positive integers, this is the irreducible finite
dimensional representation of $\mathfrak{g}$ corresponding to
the dominant  weight $\mu$.

To come back to our original  problem, consider a finite dimensional 
irreducible representation ${\cal D}_\mu$, with highest weight
$\mu = \sum \limits_{i=1}^r n_i \mu_i, n_i \in \mathbb N$. 
The basic idea to try to 
define a structure of an $F-$Lie algebra
on  $S= B \oplus A_1 =\left(\mathfrak{g} \oplus {\cal D}_\mu\right) 
\oplus {\cal D}_{\frac{\mu}{F}}$ since, roughly speaking,
the representations ${\cal D}_\mu$ and ${\cal D}_{\frac{\mu}{F}}$ can
 be related:

\begin{equation}
\label{eq:fsusy}
{\cal S}^F \left({\cal D}_{\mu/F} \right) \sim {\cal D}_\mu.
\end{equation} 

\noindent
Indeed
$|\frac{\mu}{F}>^{\otimes F} 
\in {\cal S}^F\left({\cal D}_{\frac{\mu}{F}}\right)$
and $|\mu> \in {\cal D}_\mu$
both  satisfy  Eq.[\ref{eq:hw}]. 
However,
the representation ${\cal D}_\mu$ is finite dimensional but
the sub-representation of ${\cal S}^F \left({\cal D}_{\frac{\mu}{F}}\right)$
generated by $|\frac{\mu}{F}>^{\otimes F}$ is infinite dimensional
\cite{flie,poly}.
Thus, the main difficulty, in such a 
construction, is to do with  the requirement of 
relating an infinite dimensional
representation ${\cal D}_{\mu/F}$ to a finite dimensional 
representation  ${\cal D}_\mu$ 
in an equivariant
way, {\it i.e.} respecting the action of $\mathfrak{g}$.
 One possible way of overcoming
this difficulty is to embed 
${\cal D}_\mu$ into an infinite dimensional
(reducible but indecomposable) representation \cite{flie,poly,infty}.
Another possibility is to embed $\mathfrak{g}$ into an infinite dimensional 
algebra  (dubbed $V(\mathfrak{g}))$ \cite{flie,infty, km}) and extend
the representations ${\cal D}_\mu$ and ${\cal D}_{\frac{\mu}{F}}$
to representations of $V(\mathfrak{g})$.

There is another difficulty related to such a construction. If one starts with
$\mathfrak{D}_{\mu_1}$, the vector representation of $\mathfrak{so}(1,d-1)$,
the representation $\mathfrak{D}_{\frac{\mu_1}{F}}$ cannot be exponentiated
(see \cite{kpr} {\it e.g.}) and does not define a representation of the
 Lie group
$\overline{SO(1,d-1)}$ (the universal covering group of $SO(1,d-1)$)
except when $d=3$, where such representations describe relativistic
anyons \cite{fsusy3d, anyon}.                        
\section{Example of finite dimensional $F-$Lie algebras}
In the previous section, we indicated how one can construct
infinite dimensional examples 
 of $F-$Lie algebras.  In this section, with the classification
of Lie (super)algebras as a guideline, we will give an inductive
construction of   finite dimensional $F-$Lie algebras. 

\noindent
In what follows $S$ is a  $1-$Lie algebra  means:

\begin{enumerate}
\item $S=\mathfrak{g}_0 \oplus \mathfrak{g}_1$, with
$\mathfrak{g}_0$ a Lie algebra and $\mathfrak{g}_1$ 
is a representation of $\mathfrak{g}_0$ isomorphic to the
adjoint representation;
\item there is a $\mathfrak{g}_0-$ equivariant map 
$\mu :\mathfrak{g}_1 \to \mathfrak{g}_0$ such that 
$\left[f_1, \mu(f_2) \right] + \left[f_2, \mu(f_1) \right] =0,
f_1,f_2 \in \mathfrak{g}_1$.
\end{enumerate} 
The basic result is the following theorem:

\begin{theo}
Let $\mathfrak{g_0}$ be a (complex) Lie algebra and 
$\mathfrak{g}_1$ a representation of $\mathfrak{g_0}$. Suppose given

(i) the structure of an $F_1-$Lie algebra on  
$S_1=\mathfrak{g_0} \oplus \mathfrak{g_1}$;
 
(ii)  the structure of an $F_2-$Lie algebra on  
$S_2=\mathbb C \oplus \mathfrak{g_1}$.

\noindent
Then $S=\big(\mathfrak{g_0} \otimes \mathbb C\big) \oplus \mathfrak{g_1}$ 
can be given the structure of an $(F_1+F_2)-$Lie algebra.
    
\end{theo}    

{\it Proof}

\noindent
There exists (i)  a $\mathfrak{g_0}-$equivariant map
 $\mu_1:{\cal S}^{F_1}\left(\mathfrak{g_1}\right) \longrightarrow 
\mathfrak{g}_0$  and (ii) a $\mathfrak{g_0}-$equivariant map
 $\mu_2:{\cal S}^{F_2}\left(\mathfrak{g_1}\right) \longrightarrow 
\mathbb C$, the second  map is just 
a symmetric $F_2^{\mathrm{th}}-$order invariant form on $\mathfrak{g}_1$.
Now, consider $\mu:{\cal S}^{F_1+ F_2}\left(\mathfrak{g_1}\right) 
\longrightarrow \mathfrak{g}_0 \otimes\mathbb C \cong \mathfrak{g}_0$ defined 
by $\forall f_1,\cdots,f_{F_1+F_2} \in \mathfrak{g_1}:$

\begin{eqnarray}
\label{eq:tensor}
&& \hskip 4truecm
\mu(f_1,\cdots,f_{F_1+F_2})=  \\
&&\frac{1}{F_1 !}\frac{1}{F_2 !} \sum \limits_{\sigma \in S_{F_1 + F_2}}
\mu_1(f_{\sigma(1)},\cdots,f_{\sigma({f_{F_1}})}) \otimes 
\mu_2(f_{\sigma(f_{F_1+1})},\cdots,f_{\sigma(f_{F_1+F_2})}).
\nonumber
\end{eqnarray}

\noindent
with $S_{F_1 + F_2}$ the group of permutations of $F_1 + F_2$ elements.
By construction, this map is a $\mathfrak{g_0}-$equivariant map
from ${\cal S}^{F_1+ F_2}\left(\mathfrak{g_1}\right) \longrightarrow
\mathfrak{g_0}$, 
thus the first three  Jacobi identities (\ref{eq:jac})
are clearly satisfied.  The last Jacobi identity
is more difficult to check and  is directly related to the last Jacobi
identity for the $F_1-$Lie algebra $S_1$ by  a factorisation property.
Indeed (with $F=F_1+F_2$) if one calculates:

$$
\sum\limits_{i=0}^{F} \left[ f_i,\mu\left(f_0,\dots,
f_{i-1},
f_{i+1},\dots,f_{F}\right) \right],
$$

\noindent
and selects terms of the form (with $\sigma \in S_{F_1+F_2 +1)}$)  
$
\mu_1(f_{\sigma(1)},\cdots,f_{\sigma({f_{F_1}})}) \otimes 
\mu_2(f_{\sigma(f_{F_1+1})},\cdots,f_{\sigma(f_{F_1+F_2})}$,
using $\mu_2(f_{\sigma(f_{F_1+1})},\cdots,f_{\sigma(f_{F_1+F_2})}
 \in \mathbb C$
the identity reduces to

\begin{eqnarray}
&&\sum\limits_{i=0}^{F_1} 
 \left[ f_{\sigma(i)},\mu_1\left(f_{\sigma(1)},\cdots,
f_{\sigma(i-1)},f_{\sigma(i+1)},\cdots,f_{\sigma({f_{F_1}})}
\right) \right] \nonumber \\
&&\otimes 
\mu_2(f_{\sigma(f_{F_1+1})},\cdots,f_{\sigma(f_{F_1+F_2})}=0.
\nonumber
\end{eqnarray}

\noindent
This follows from the corresponding Jacobi identity for
the $F_1-$Lie algebra $S_1$. Now proceeding along the same lines for the other
terms, a similar factorisation works. Thus the fourth Jacobi identity
is satisfied and $S$ is an $(F_1+F_2)-$Lie algebra.

{\it QED} \\

Here there are some families of examples:
\begin{enumerate}
\item $S_1= \mathfrak{g}\oplus \mathrm {Ad} (\mathfrak{g})$( a $1-$Lie
algebra);

\noindent
$S_2= \mathbb C \oplus \mathrm {Ad}(\mathfrak{g})$ 
(a Lie superalgebra if $\mathfrak{g}$ admits an equivariant quadratic form).

\item If $\mathfrak{g} = \mathfrak{g}_0 \oplus \mathfrak{g}_1$
is a Lie superalgebra (basic of type I or II or $Q(n)$ \cite{fss})
we associate to $\mathfrak{g}$ an ``augmented'' Lie superalgebra as follows:

\begin{eqnarray}
\label{eq:super-F}
\mathfrak{g} = \mathfrak{g}_0 \oplus \mathfrak{g}_1 
\longrightarrow \left \{
\begin{array}{ll}
S={\cal B} \oplus {\cal F} =  \mathfrak{g}_0 \oplus \mathfrak{g}_1&
\mathrm{~if~} \mathfrak{g} \mathrm{~is~of~ type~ I } \cr
S={\cal B} \oplus {\cal F} =  \mathfrak{g}_0 \oplus
( \mathfrak{g}_1 \oplus\mathfrak{g}_1)&
\mathrm{~if~} \mathfrak{g} \mathrm{~is~of~ type~ II }  \cr
S={\cal B} \oplus {\cal F} =  \mathfrak{g}_0 \oplus \mathfrak{g}_1&
\mathrm{~if~} \mathfrak{g}= Q(n),
\end{array}
\right.
\end{eqnarray}

\noindent
The non-zero graded part of these ``augmented'' Lie superalgebras 
always admits a $\mathfrak{g}_0$ invariant quadratic form and hence
$S_2=\mathbb C \oplus {\cal F}$ is a Lie superalgebra:

For the type I superalgebras  we have 
$\mathfrak{g}_1 = \mathfrak{D} \oplus \mathfrak{D}^*$ (see \cite{fss}), 
and so one has a natural map:
${\cal S}^2  \left(\mathfrak{D} \oplus \mathfrak{D}^*\right) 
\longrightarrow \mathbb C$.

For the type II superalgebras, we recall  that $\mathfrak{g}_1$
admits an invariant antisymmetric bilinear form and hence 
$\mathfrak{g}_1 = {\cal D}$ is self-dual \cite{fss}. Therefore,
there is an invariant quadratic form on ${\cal F}=
\mathfrak{g}_1\oplus \mathfrak{g}_1$.

For the strange superalgebra $Q(n)$, $\mathfrak{g}_0=\mathfrak{sl}(n+1)$, 
  the representation
$\mathfrak{g}_1$ is the adjoint representation of $\mathfrak{g}_0$
(see \cite{fss}) and hence admits an invariant quadratic form 
(the Killing form). \\

The existence of an invariant bilinear form on $\mathfrak{g}_1$ 
({\it i.e.} before the ``augmentation'' (\ref{eq:super-F}))
means that there is a 
$\mathfrak{g}_0-$equivariant
mapping  ${\cal S_\pm}^2(\mathfrak{g}_1) \longrightarrow \mathbb C$ 
(where ${\cal S_+}^2(\mathfrak{g}_1)$
(resp. ${\cal S_-}^2(\mathfrak{g}_1)$) represent the two-fold symmetric 
(antisymmetric)
tensor product of $\mathfrak{g}_1$).
We denote generically this tensor by $\delta_{\alpha \beta}$ when it
is symmetric and $\Omega_{\alpha \beta}$ when it is antisymmetric.
This  can   equivalently  be rewritten in a basis of $\mathfrak{g}_1,
F_\alpha \in \mathfrak{g}_1$

\begin{eqnarray}
\label{eq:super-inv}
\begin{array}{ll}
\left\{F_\alpha, F_\beta \right\} = \delta_{\alpha \beta},
&\mathrm{~for~the~type~I~superalgebras~and~for~} Q(n) \cr
\left[F_\alpha, F_\beta \right] = \Omega_{\alpha \beta},
& \mathrm{~for~the~type~II~superalgebras~}. 
\end{array}
\end{eqnarray}
\noindent
with $\left\{\ ,\ \right\}$ (resp. $\left[\ ,\ \right]$) the symmetric 
(resp. antisymmetric) bilinear forms.

However, after the ``augmentation ''(\ref{eq:super-F}) 
in the case of Lie superalgebra  of type II,
the mapping ${\cal S}^2\left({\cal F}\right) \longrightarrow \mathbb C$
({\it i.e.} the quadratic form on ${\cal F}$) reads

\begin{eqnarray}
\label{eq:odd}
\left\{F_{i \alpha}, F_{j \beta} \right\} = \varepsilon_{ij} \Omega_{\alpha
\beta},
\end{eqnarray}

\noindent
with $F_{i\alpha} \in \mathfrak{g}_1 \oplus \mathfrak{g}_1$.
The index $\alpha$ represents the $\mathfrak{g}_1$ degrees of freedom,
the index $i( i=-1,1)$ the two copies of 
$\mathfrak{g}_1$ and $\varepsilon_{ij}$ the  two
dimensional antisymmetric tensor.\\
\end{enumerate}

To conclude, we will give an  example of a $3-$Lie (resp.
$4-$Lie) algebra, associated
to a $1-$Lie  algebra (resp. superalgebra). 

   
\begin{ex}
Let $\mathfrak{g_0}$ be a Lie algebra and $\mathfrak{g_1}$ the
adjoint representation of $\mathfrak{g_0}$ and 
$S_3=\mathfrak{g_0} \oplus \mathfrak{g_1}$. 
We introduce
$J_a, A_a, a=1,\cdots, $ $\mathrm{dim}(\mathfrak{g}_0)$ a basis of $S_{3}$.
We denote $\mathrm{tr}({A_a A_b})= g_{ab}$ the Killing form.   
The trilinear bracket of the $3-$Lie algebra $S_3$,
associated to the Lie algebra $\mathfrak{g}$, is:

\begin{eqnarray}
\label{eq:3-lie}
\left\{A_a, A_b, A_c \right\} =
g_{ab} J_c + g_{ac} J_b + g_{bc} J_a.
\end{eqnarray}

\end{ex}  
  
If $\mathfrak{g}=\mathfrak{sl}(2)$, this is the $3-$Lie algebra
constructed in   \cite{ayu}. 

\begin{ex}
As a second example we give the formulae for the
quadrilinear bracket of the 
$4-$Lie algebra constructed  from the orthosymplectic superalgebra.
Starting from $\mathfrak{osp}(m|2n)=
\Big( \mathfrak{so}(m) \oplus \mathfrak{sp}(2n)\Big)
\oplus(\mathbf{m}, \mathbf{2n}) $, we define 
$
\mathfrak{osp}(m|2n;4)= 
\Big( \mathfrak{so}(m) \oplus \mathfrak{sp}(2n) \oplus \mathfrak{u}(1)\Big)
\oplus \Big( (\mathbf{m}, \mathbf{2n})^+ \oplus (\mathbf{m}, \mathbf{2n})^-
\Big).
$

\noindent
Let $F_{q i \alpha}$
($q=-1,+1$, $1 \le i \le m, 1 \le \alpha \le 2n$) 
denote the odd part, $J_{i j}$ the $\mathfrak{so}(m)$
generators, $S_{\alpha \beta}$ the $\mathfrak{sp}(2n)$ generators
and $h$ the $\mathfrak{u}(1)$ generator
($J_{ij}$ are antisymmetric and $S_{\alpha \beta}$ are symmetric).
The invariant tensor on $\mathfrak{so}(m)$ is given by
the symmetric tensor $\delta_{ij}$ and on $\mathfrak{sp}(2n)$
by the antisymmetric tensor $\Omega_{\alpha \beta}$, hence the invariant
tensor for $\mathfrak{osp}(m|2n)$ is  given by $\delta_{ij} 
\Omega_{\alpha \beta}$. Thus,  the quadrilinear bracket of
the $4-$algebra  takes the form

\begin{eqnarray}
\label{eq:flie-orth}
&&\left\{F_{q_1 i_1 \alpha_1}, F_{q_2 i_2 \alpha_2}, F_{q_3 i_3 \alpha_3},
F_{q_4 i_4 \alpha_4} \right\}=  \\
&&\varepsilon_{q_1 q_2}  \delta_{i_1 i_2}\Omega_{\alpha_1  \alpha_2}
\left(\delta_{q_3 + q_4} \delta_{i_3 + i_4} S_{\alpha_3 \alpha_4} + 
\delta_{q_3 + q_4} \Omega_{\alpha_3 \alpha_4} J_{i_3 i_4} 
+a \delta_{q_3 + q_4}\varepsilon_{i_3  i_4} \Omega_{\alpha_3 \alpha_4} h
\right) \nonumber  \\
&+&
\varepsilon_{q_1 q_3} \delta_{i_1 i_3} \Omega_{\alpha_1  \alpha_3}
\left(\delta_{q_2 + q_4} \delta_{i_2 + i_4} S_{\alpha_2 \alpha_4} + 
\delta_{q_2 + q_4} \Omega_{\alpha_2 \alpha_4} J_{i_2 i_4} 
+a \delta_{q_2 + q_4}\varepsilon_{i_2  i_4}  \Omega_{\alpha_2 \alpha_4} h
\right) \nonumber \\
&+ &
\varepsilon_{q_1 q_4}  \delta_{i_1 i_4}\Omega_{\alpha_1  \alpha_4}
\left(\delta_{q_2 + q_3} \delta_{i_2 + i_3} S_{\alpha_2 \alpha_3} 
+ \delta_{q_2 + q_3} \Omega_{\alpha_2 \alpha_3} J_{i_2 i_3} 
+ a \delta_{q_2 + q_3}\varepsilon_{i_2  i_3} \Omega_{\alpha_2 \alpha_3} h
\right)\nonumber  \\
& +&  
\varepsilon_{q_2 q_3} \delta_{i_2 i_3} \Omega_{\alpha_2  \alpha_3}
\left( \delta_{q_1 + q_4} \delta_{i_1 + i_4} S_{\alpha_1 \alpha_4} + 
 \delta_{q_1 + q_4} \Omega_{\alpha_1 \alpha_4} J_{i_1 i_4} 
+a \delta_{q_1 + q_4}\varepsilon_{i_1  i_4} \Omega_{\alpha_1 \alpha_4} h
\right) \nonumber \\
&+&
\varepsilon_{q_2 q_4} \delta_{i_2 i_4} \Omega_{\alpha_2  \alpha_4}
\left(\delta_{q_1 + q_3} \delta_{i_1 + i_3} S_{\alpha_1 \alpha_3} + 
 \delta_{q_1 + q_3}\Omega_{\alpha_1 \alpha_3} J_{i_1 i_3} 
+a \delta_{q_1 + q_3}\varepsilon_{i_1  i_3} \Omega_{\alpha_1 \alpha_3} h
\right) \nonumber \\
&+&
\varepsilon_{q_3 q_4}  \delta_{i_3 i_4}\Omega_{\alpha_3  \alpha_4}
\left(\delta_{q_1 + q_2} \delta_{i_1 + i_2} S_{\alpha_1 \alpha_2} + 
\delta_{q_1 + q_2} \Omega_{\alpha_1 \alpha_2} J_{i_1 i_2} 
+a \delta_{q_1 + q_2}\varepsilon_{i_1  i_2} \Omega_{\alpha_1 \alpha_2} h
\right), \nonumber
\end{eqnarray}

\noindent
with $a \in \mathbb C$. 
\end{ex}

\begin{rema}
It should be noticed that $F-$Lie algebras associated
to Lie algebras (resp. to Lie superalgebras) are of
odd (resp. even) order.
\end{rema}      

\section{Conclusion}
In this paper a sketch of the construction of $F-$Lie 
algebras associated to Lie (super)algebras was given. More complete
results, such as 
 a criteria for simplicity, 
representation theory, matrix realisations {\it etc.}, will be given elswhere.

\subsection*{Acknowledgements}        
J.~Lukierski is gratefully acknowledged for useful discussions and remarks.


\vskip 1.truecm    
\begin{thebibliography}{99}
\footnotesize
%
\bibitem{cm}
S.~Coleman and J.~Mandula, {\it Phys. Rev.} {\bf 159} (1967) 1251.
%
\bibitem{hls}
R.~Haag, J.~T.~Lopuszanski and M.~F.~Sohnius, {\it Nucl. Phys.} {\bf B88} 
(1975) 257.
%
\bibitem{ker}
R.~Kerner, J. Math. Phys. {\bf 33} (1992) 403-411,
%
R.~Kerner, Class. Quantum Grav. {\bf 9} (1992) S137.
%
\bibitem{luis}
L.~A.~Wills Toro, J. Math. Phys. {\bf 36} (1995) 2085.
%
\bibitem{para}
V.~A.~ Rubakov, V.~P.~ Spiridonov, {\it Mod. Phys. Lett. } {\bf A 3} 
(1988) 1337;
%
A. Khare, {\it J. Math. Phys. } {\bf 34} (1993) 1277;
%
R.~Floreanini, L.~Vinet, {\it Phys. Rev.} {\bf D44} (1991) 3851;
%
J~. Beckers, N.~ Debergh, {\it Int. J. Mod. Phys. } {\bf A 8} (1993), 5041;
%
A.~G.~ Nikitin, V.~V.~ Tretynyk,  {\it J. Phys. } {\bf A 28} (1995) 1655. 
%
\bibitem{fsusy}
C.~Ahn, D.~Bernard and A.~ Leclair, Nucl. Phys. {\bf B346} (1990) 409.
%
\bibitem{fsusy1d}
J.~ L.~ Matheus-Valle and Marco A.~ R.~ Monteiro, Mod. Phys. Lett. {\bf A7}
(1992) 3023;
%
S.~ Durand, Mod. Phys. Lett {\bf A8}  (1993) 2323 [hep-th/9305130];
%
N.~ Debergh, J. Phys. {\bf A26} (1993) 7219;
%
N.~ Mohammedi,  Mod. Phys. Lett. { \bf A10} (1995) 1287 [hep-th/9412133];
%
L.~ P.~ Colatto and J.~ L.~ Matheus-Valle, J. Math. Phys. {\bf 37} (1996)
6121 [hep-th/9504101].
%
%
\bibitem{fr}
N.~ Fleury and M. Rausch de Traubenberg, Mod. Phys. Lett. {\bf A11} (1996)
899 [hep-th/9510108].
%
\bibitem{qfsusy}
R.~S.~Dunne, A.~J.~Macfarlane, J.~A.~de Azcarraga and J.~C.~Perez Bueno,
{\it  Int. J.  Mod. Phys. } {\bf A 12} (1997) 3275;
[hep-th/9610087];
%
R.~S.~Dunne,
hep-th/9703137;
%
M.~Daoud, Y.~Hassouni and M.~Kibler,
in Symmetries in Science X, eds.
B.Gruber and M.Ramek (Plenum, New York,1998) [quant-ph/9710016];
%
M.~Daoud, Y.~Hassouni and M.~Kibler,
{\it Phys.\ Atom.\ Nucl.\ } {\bf 61} (1998) 1821,
[{\it Yad.\ Fiz.\ }  {\bf 61} (1998) 1935]
[quant-ph/9804046];
%
M.~Daoud and M.~Kibler,
math-ph/9912024;
%
M.~Mansour, M.~ Daoud, Y.~ Hassouni, 
{\it  Rep. Math. Phys.} {\bf  44} (1999) 435;
M.~Mansour, M.~ Daoud, Y.~ Hassouni, 
{\it  Phys.\ Lett.\ }  {\bf B  454} (1999) 281.
%
\bibitem{hek}
E.~H.~El Kinani,
{\it Mod.\ Phys.\ Lett.\ }   {\bf A15} (2000) 2139.
%
\bibitem{asm}
K.~Aghababaei Samani and A.~Mostafazadeh,
{\it Nucl.\ Phys.\ }  {\bf B595} (2001) 467
[hep-th/0007008].
%
\bibitem{am}
J.~A.~ de Azc\`arraga and A.~J.~ Macfarlane, J. Math .Phys. {\bf 37}
(1996) 1115 [hep-th/9506177].
%
\bibitem{prs}
A.~Perez, M. Rausch de Traubenberg and P. Simon
 Nucl. Phys. {\bf B482} (1996), 325 [hep-th/9603149];
%
M. Rausch de Traubenberg and P. Simon, Nucl. Phys. {\bf B517} (1998) 485
[hep-th/9606188].
%
\bibitem{fsusy2d}
J.~ L.~Matheus-Valle and Marco A.~ R.~Monteiro,  Phys. Lett. {\bf B300}
(1993) 66;
%
E.~H.~Saidi, M.~B.~Sedra and J.~Zerouaoui, Class. and
Quantum Grav.  {\bf 12} (1995) 1567;
%
E.~H.~Saidi, M.~B.~Sedra and J.~Zerouaoui, Class.Quant.Grav.{
 \bf 12} (1995) 2705.
%
\bibitem{ad}
H.~Ahmedov and O.~F.~Dayi,
{\it Mod.\ Phys.\ Lett.\ }  {\bf A15} (2000) 1801
[math.qa/9905164];
H.~Ahmedov and O.~F.~Dayi,
{\it J.\ Phys.\ } {\bf A32} (1999) 6247
[math.qa/9903093].
%
\bibitem{kk}
F.~Kheirandish and M.~Khorrami,
{\it Eur.\ Phys.\ J.\ } {\bf C18} (2001) 795
[hep-th/0007013];
F.~Kheirandish and M.~Khorrami,
{\it   Int. J .\ Mod.\ Phys.}\ {\bf A 16} (2001) 2165
[hep-th/0004154].
%
\bibitem{fvir}
S.~ Durand, Mod. Phys. Lett. {\bf A7}  (1992) 2905 [hep-th/9205086].
%
\bibitem{fsusy3d}
M.~Rausch de Traubenberg and M.~Slupinski,
Mod. Phys. Lett. {\bf A12} (1997) 3051 [hep-th/9609203].
%
\bibitem{brs}
I.~Benkaddour, A.~El Rhalami and E.~H.~Saidi,
 hep-th/0007142;
I.~Benkaddour, A.~El Rhalami and E.~H.~Saidi,
hep-th/0101188.
%
\bibitem{fsusyh}
M.~Rausch de Traubenberg, hep-th/9802141 (Habilitation Thesis, in French).
%
\bibitem{flie}
M.~Rausch de Traubenberg and M.~J.~Slupinski, {\it J. Math. Phys} {\bf  41}
(2000) 4556 [hep-th/9904126].
%
%
\bibitem{poly}
M. Rausch de Traubenberg, Proceedings of Workshop on High Energy Physics 2 
(2000) 19,  [hep-th/0007150].
%
\bibitem{infty}
M.Rausch de Traubenberg, { \it Nucl. Phys. Proc. Suppl. } {\bf 102 } 
 (2001) 256 [hep-th/0109106].
\bibitem{km}
M. Rausch de Traubenberg and M. Slupinski,
Preprint PM/01-23, math.RT/0109090.
%
%
\bibitem{kpr}
S.~M.~Klishevich, M~S.~Plyushchay and M.~Rausch de Traubenberg,
hep-th/0101190, to appear in {\it Nucl. Phys.} {\bf B}.
%
\bibitem{anyon}
M.~S~. Plyushchay, {\it Phys. Lett.} {\bf B273} (1991) 250;
%
R.~Jackiw and V.~P.~ Nair, {\it Phys. Rev.} {\bf D43} (1991) 1933.
%
\bibitem{fss}
L.~Frappat, P.~Sorba and  A.~Sciarrino, hep-th/9607161.
%
\bibitem{ayu}
H.~Ahmedov, A.~Yildiz and  Y. Ucan, {\it J.\ Phys. }  {\bf A 34} (2001) 6413 
[math.rt/0012058].
%
\end{thebibliography}
\end{document}